\documentclass[sigconf]{acmart}

\usepackage{balance}

\def\BibTeX{{\rm B\kern-.05em{\sc i\kern-.025em b}\kern-.08emT\kern-.1667em\lower.7ex\hbox{E}\kern-.125emX}}
    
\copyrightyear{2020} 
\acmYear{2020} 
\setcopyright{acmlicensed}\acmConference[ICMR '20]{Proceedings of the 2020 International Conference on Multimedia Retrieval}{October 26--29, 2020}{Dublin, Ireland}
\acmBooktitle{Proceedings of the 2020 International Conference on Multimedia Retrieval (ICMR '20), October 26--29, 2020, Dublin, Ireland}
\acmPrice{15.00}
\acmDOI{10.1145/3372278.3391927}
\acmISBN{978-1-4503-7087-5/20/06}

\begin{document}

\fancyhead{}

\title{Automatic Reminiscence Therapy for Dementia}

\author{Mariona Carós}
\affiliation{%
 \institution{Universitat Politecnica de Catalunya}}

\author{Maite Garolera}
\affiliation{\institution{Consorci Sanitari de Terrassa}}
\email{mgarolera@cst.cat}

\author{Petia Radeva}
\affiliation{\institution{Universitat de Barcelona}}
\email{petia.ivanova@ub.edu}

\author{Xavier Giro-i-Nieto}
\affiliation{\institution{Universitat Politecnica de Catalunya}}
\email{xavier.giro@upc.edu}

\renewcommand{\shortauthors}{Carós, et al.}

%
\begin{abstract}
 With people living longer than ever, the number of cases with dementia such as Alzheimer's disease increases steadily. It affects more than 46 million people worldwide, and it is estimated that in 2050 more than 100 million will be affected. While there are no effective treatments for these terminal diseases, therapies such as reminiscence, that stimulate memories from the past are recommended. Currently, reminiscence therapy takes place in care homes and is guided by a therapist or a carer. In this work, we present an AI-based solution to automate the reminiscence therapy. This consists of a dialogue system that uses photos of the users as input to generate questions about their life. Overall, this paper presents how reminiscence therapy can be automated by using deep learning, and deployed to smartphones and laptops, making the therapy more accessible to every person affected by dementia.
\end{abstract}

%
%

\ccsdesc[500]{Computing methodologies~Natural language generation}
\ccsdesc[500]{Computing methodologies~Neural networks}
\ccsdesc[500]{Applied computing~Consumer health}

%
\keywords{Reminiscence therapy, Alzheimer, dementia, generative dialogue system, chatbot, visual question generator}
%

%
\maketitle

\section{Introduction}
Increases in life expectancy in the last century have resulted in a large number of people living to old ages and will result in a double number of dementia cases by the middle of the century \cite{hickman2016alzheimer}\cite{olanrewaju2015multimodal}. The most common form of dementia is Alzheimer disease which contributes to 60–70\% of cases \cite{jindal2014alzheimer}. Research focused on identifying treatments to slow down the evolution of Alzheimer's disease is a very active pursuit, but it has been only successful in terms of developing therapies that eases the symptoms without addressing the cause \cite{alzheimer20152015}\cite{solas2015treatment}.
Furthermore, people with dementia might have some barriers to access these therapies, such as cost, availability and displacement to the care home or hospital, where the therapy takes place. We believe that Artificial Intelligence (AI) can contribute in innovative systems to give accessibility and offer new solutions to the patients needs, as well as help relatives and caregivers to understand the illness of their family member or patient and monitor the progress of the dementia.

Therapies such as reminiscence, that stimulate memories of the patient's past, have well documented benefits on social, mental and emotional well-being \cite{subramaniam2012impact}\cite{huldtgren2015reminiscence}, making them a very desirable practice, especially for older adults. Reminiscence therapy in particular involves the discussion of events and past experiences using tangible prompts such as pictures or music to evoke memories and stimulate conversation \cite{woods2018reminiscence}. With this aim, we explore multi-modal deep learning architectures to be used to develop an intuitive, easy to use, and robust dialogue system to automate the reminiscence therapy for people affected by mild cognitive impairment or at early stages of Alzheimer's disease. 

We propose a conversational agent that simulates a reminiscence therapist by asking questions about the patient's experiences. Questions are generated from pictures provided by the patient, which contain significant moments or important people in user's life. The proposed methodology is specific for dementia therapy, compared to a general Image-based Question and Answering (Q\&A) system as \cite{yang2016stacked}, because the generated questions cannot be answered by only looking at the picture as common Q\&A systems do. The user needs to know the place, time, people or animals appearing in the picture to be able to answer the questions. To engage the user in the conversation, we propose a second model which generates comments on user's answers. A chatbot model trained with a dataset containing simple conversations between different people. The activity pretends to be challenging for the patient, as the questions may require the user to exercise the memory. However, it also intends to be amusing at the same time. Our contributions include:
\begin{itemize}
    \item Automation of the Reminiscence therapy by using a multi-modal approach that generates questions from pictures, without the need of a reminiscence therapy dataset.
    \item An end-to-end deep learning approach ready to be used by mild cognitive impairment patients. The system is designed to be intuitive and easy to use for the users and could be reached by any smartphone with internet connection.
\end{itemize}

\section{Related Work}

The origin of chatbots goes back to 1966 with the creation of ELIZA \cite{weizenbaum1966eliza} by Joseph Weizenbaum at MIT. Its implementation consisted of pattern matching and substitution methodology. Recently, data driven approaches have drawn significant attention. Existing work along this line includes retrieval-based methods \cite{ji2014information}\cite{wu2016sequential} and generation-based methods\cite{serban2016building}\cite{serban2017hierarchical}. In this work we focus on generative models, where sequence-to-sequence algorithms that use RNNs to encode and decode inputs into responses is a current best practice.
 
Our conversational agent uses two architectures to simulate a specialized reminiscence therapist. The block in charge of generating questions is based on the work \textit{Show, Attend and Tell} \cite{xu2015show}. This work generates descriptions from pictures, also known as image captioning. In our case, we focus on generating questions from pictures involving the user's life. Our second architecture is inspired by \textit{Neural Conversational Model} from \cite{vinyals2015neural} where the author presents an end-to-end approach to generate simple conversations. Building an open-domain conversational agent is a challenging problem. As addressed in \cite{zhang2018personalizing} and \cite{ghandeharioun2019approximating}, the lack of a consistent personality and lack of long-term memory which produces some meaningless responses in these models are still unresolved problems.

Regarding works related to multi-modal conversational agents, we find Visual Dialog \cite{das2017visual} where the conversational agent is the one that has to answer the questions about the image. Some works have proposed conversational agents for older adults with a variety of uses, such as stimulate conversation \cite{yasuda2013development} , palliative care \cite{utami2017talk} or daily assistance like ‘Billie’ reported in \cite{kopp2018conversational} which is a virtual agent that uses facial expression for a more natural behavior and is focused on managing the user’s calendar. These works perform well on their specific tasks, but non of them include reminiscence therapy. There is a work focusing on the content used in Reminiscence therapy \cite{bermingham2013automatically}, where the authors propose a system that recommends multimedia content to be used in therapy. To the best of our knowledge, there is no equivalent approach to the one proposed in the field. 


\section{Methodology}
\label{sec:methodology}

\begin{figure}[t]
    \centering
    \includegraphics[width=\linewidth]{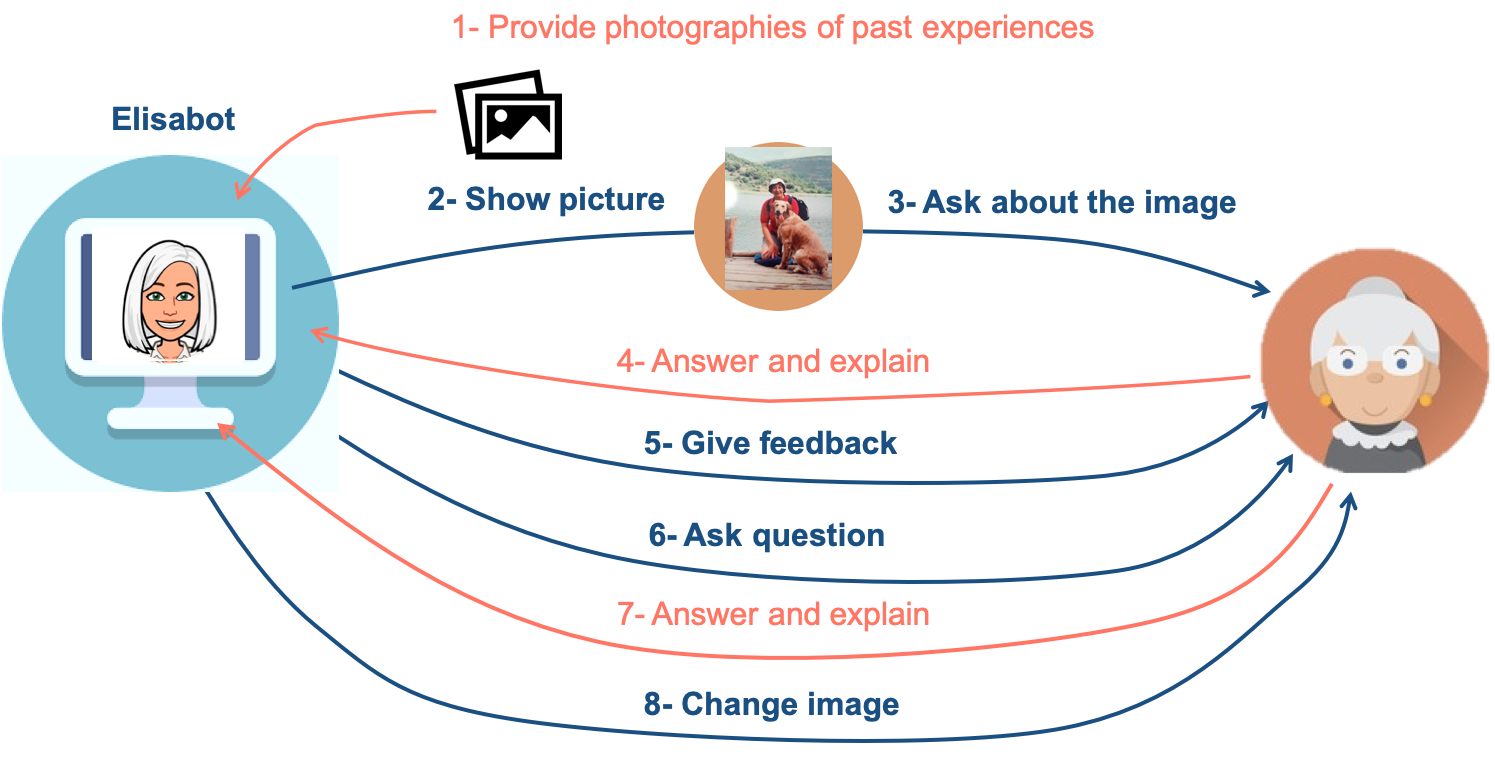}
    \caption{Scheme of the interaction with Elisabot}
    \label{fig:engine_diagram}
\end{figure}

In this section we explain how the interaction with the model works, the main two components of our model and the chosen hyperparameters that give the best performance. 

We named the model Elisabot and its goal is to maintain a dialog with the patient about their life experiences. Before starting the conversation, the user must introduce photos containing significant moments for them. The system randomly chooses one of these pictures and analyses the content. Elisabot then shows the selected picture and starts the conversation by asking a question about the picture. The user should give an answer, even though he does not know it, and Elisabot makes a relevant comment on it. The cycle starts again by asking another relevant question about the image and the flow is repeated for 4 to 6 times until the picture is changed. The Figure \ref{fig:engine_diagram} summarizes the workflow of our system.

Elisabot is composed of two models: the model in charge of asking questions about the image which we will refer to it as Visual Question Generator (VQG), and the Chatbot model which tries to make the dialogue more engaging by giving feedback to the user's answers. Both models are trained using Stochastic Gradient Descent with ADAM optimization \cite{kingma2014adam} and a learning rate of 1e-4. Furthermore, we use dropout regularization \cite{srivastava2014dropout} which prevents from over-fitting. 


\subsection{VQG model}

\begin{figure}[b]
    \centering
    \includegraphics[width=\linewidth]{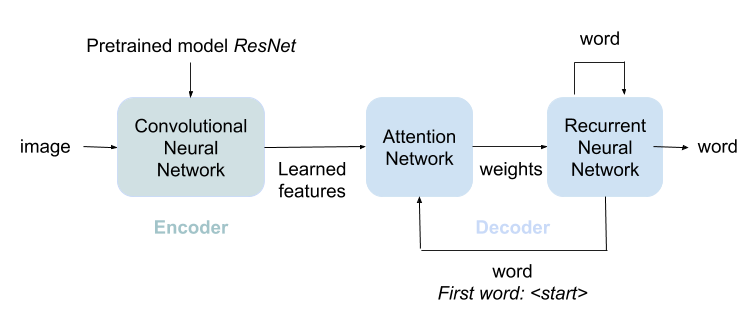}
    \caption{VQG model}
    \label{fig:vqg_model}
\end{figure}

The algorithm behind VQG consists of an Encoder-Decoder architecture with attention, as shown in Figure \ref{fig:vqg_model}. The model is trained to maximize the likelihood of producing a target sequence of words optimizing the cross-entropy loss \cite{bishop2006pattern}. The Encoder takes as input one of the given photos from the user and learns its information using a Convolutional Neural Network (CNN). The CNN provides the image's learned features to the Decoder which generates the question word by word by using an attention mechanism with a Long Short-Term Memory (LSTM). Since there are already CNNs trained on large datasets with an outstanding performance, we integrate a \textit{ResNet-101} \cite{he2016deep} trained on ImageNet. 

 Regarding hyperparameters, the VQG encoder is composed of 2048 neuron cells, while the VQG decoder has an attention layer of 512 followed by an embedding layer of 512 and a LSTM with the same size. We set the batch size to 32. We use a dropout of 50\% and a beam search of 7 for decoding, which let as obtain up to 5 output questions. The vocabulary we use consists of all words seen 3 or more times in the training set, which amounts to 11.214 unique tokens. Unknown words are mapped to an $<$\texttt{unk}$>$ token during training, but we do not allow the decoder to produce this token at test time. We also set a maximum sequence length of 6 words as we want simple questions that are easy to understand and easy to learn by the model.

\subsection{Chatbot network}
\label{sec:chatbot}

\begin{figure}[t]
    \centering
    \includegraphics[width=\linewidth]{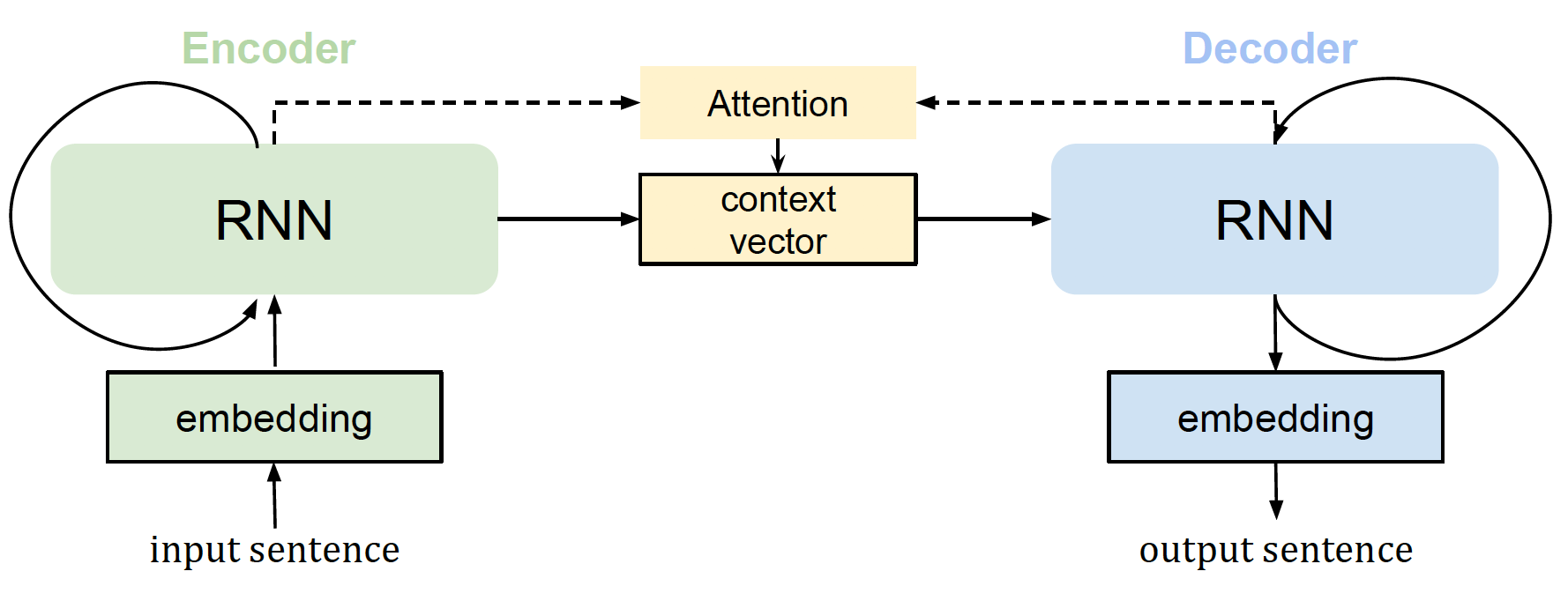}
    \caption{Chatbot model}
    \label{fig:seq-to-seq}
\end{figure}

The core of our chatbot model is a sequence-to-sequence \cite{sutskever2014sequence}, which is shown in Figure \ref{fig:seq-to-seq}. The encoder iterates through the input sentence one word at each time step producing an output vector and a hidden state vector. The hidden state vector is passed to the next time step, while the output vector is stored. We use a bidirectional Gated Recurrent Unit (GRU), one GRU fed in sequential order and another one fed in reverse order. The outputs of both networks are summed at each time step, so we encode past and future context. 

By using an attention mechanism, the decoder uses the encoder’s context vectors, and internal hidden states to generate the next word in the sequence. It continues generating words until it outputs an $<$\texttt{end}$>$ token. We use an attention layer to multiply attention weights to encoder's outputs to focus on the relevant information when decoding the sequence. This approach has shown better performance on sequence-to-sequence models \cite{bahdanau2014neural}.

In this model we use a hidden size of 500 and Dropout regularization of 25\%. We set the batch size to 64. We use greedy search for decoding, which consists of making the optimal token choice at each step. We first train it with Persona-chat and then fine-tune it with Cornell dataset. The vocabulary we use consists of all words seen 3 or more times in Persona-chat dataset and we set a maximum sequence length of 12 words.


\section{Datasets}

The lack of open-source datasets containing dialogues from reminiscence therapy lead us to use the following public datasets: a dataset that maps pictures with questions and an open-domain conversation dataset. The details are as follows.

\subsection{MS-COCO, Bing and Flickr datasets}

We use MS COCO, Bing and Flickr datasets provided by \cite{naturalVGQ} to train the model that generates questions. These datasets contain natural questions about images with the purpose of knowing more about the picture. Questions cannot be answered by only looking at the image. Each source contains 5,000 images with 5 questions per image, adding a total of 15,000 images with 75,000 questions. COCO dataset includes images of complex everyday scenes containing common objects in their natural context, but it is limited in terms of the concepts it covers. Bing dataset contains more event related questions and has a wider range of questions longitudes (between 3 and 20 words), while Flickr questions are shorter (less than 6 words) and the images appear to be more casual. We use 80\% of data for training, 10\% for validation and 10\% for testing.

\subsection{Persona-chat and Cornell-movie corpus}

We use two datasets to train our chatbot model. The first one is the Persona-chat \cite{zhang2018personalizing} which contains dialogues between two people  with different profiles that are trying to know each other. It is complemented by the Cornell-movie dialogues dataset \cite{Danescu_Cornell}, which contains a collection of fictional conversations extracted from raw movie scripts. Persona-chat's sentences have a maximum of 15 words, making it easier to learn for machines and a total of 162,064 utterances over 10,907 dialogues. Cornell-movie dataset contains 304,713 utterances over 220,579 conversational exchanges between 10,292 pairs of movie characters.

\section{Validation}
\label{sec:training}

An important aspect of dialogue response generation systems is how to evaluate the quality of the generated response. This section presents the validation procedure together with some qualitative results.

\subsection{Validation procedure}

Our first block of the system, the VQG model, is validated with BLEU score \cite{papineni2002bleu}, which is a measure of similitude between generated and target sequences of words, widely used in natural language processing. It assumes that valid generated responses have significant word overlap with the ground truth responses. 

Our chatbot model instead, only has one reference ground truth in training when generating a sequence of words. We consider that BLEU is not a good metric to check, as in some occasions responses have the same meaning, but do not share any words in common. Thus, we save several models with different hyperparameters and at different number of training iterations. Then, we compare them using human evaluation to chose the model that performs better in a conversation.

\subsection{Qualitative results}
\label{sec:qualitative}

\begin{table}[b]
  \caption{Generated questions}
  \label{tab:gen_questions}
  \centering
  \begin{tabular}{|l|p{0.6\linewidth}|}
    \toprule
    \cmidrule(r){1-2}
    Input photo     & Generated questions \\
    \midrule
    \raisebox{-\totalheight}{\includegraphics[width=0.3\linewidth]{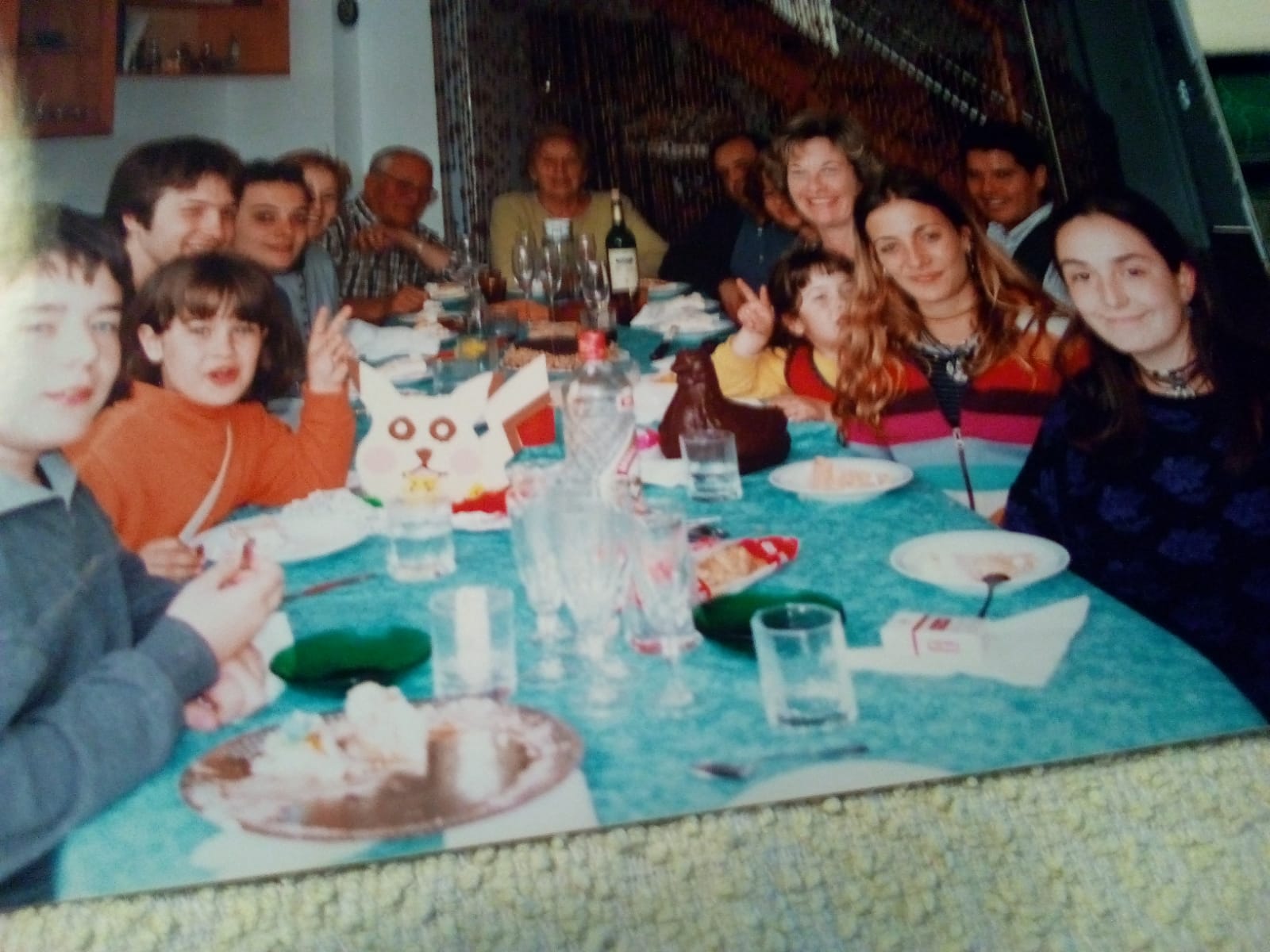}} 
    & 
    \begin{itemize}
    \item What kind of cake is that?
    \item Who made the cake?
    \item Is this a birthday cake?
    \item How old is the birthday person?
    \end{itemize}
    \\
    \midrule
    
    \raisebox{-\totalheight}{\includegraphics[width=0.3\linewidth]{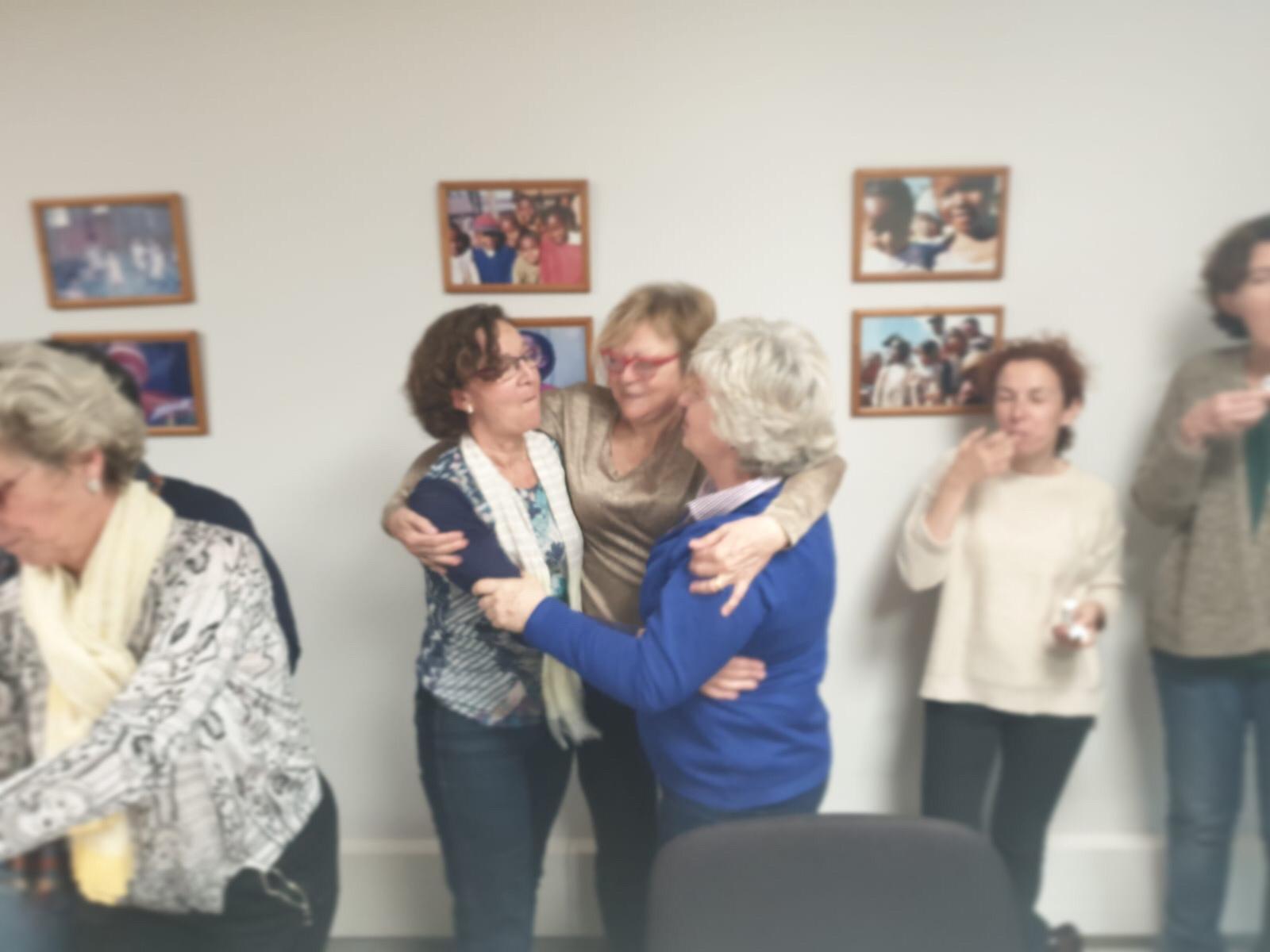}}
    &
    \begin{itemize}
    \item What kind of party is that?
    \item Why are they holding you up?
    \item What is the man looking at?
    \end{itemize}
    \\
    \bottomrule
  \end{tabular}
\end{table}

Table \ref{tab:gen_questions} contains several generated questions by our VQG model for two of the proposed images, taken from the user's photo albums. By looking at the questions, we can easily realise that the model is not only able to generate questions grammatically correct, but to understand the content of the image and ask about it. Most of the generated questions could be asked by a person. Nonetheless, we found some of them such as \textit{What is the man looking at?} that do not correspond with the image's content.

The following lines show the behavior of the chatbot model, together with the generated questions, for the second image of Table \ref{tab:gen_questions}. The dialogue is a sample of the usability study with a patient, where each sentence starting with "$>$" is an input to our seq2seq model. We made the first question asked by the model to be always time related to place the user in time, as it was advised by the doctor involved in this work. In the experiment we see that in some occasions the feedback given by Elisabot do not correspond with the context of the dialogue, however most of the time is right and even funny. We provide a demo video of this work at \cite{webElisabot}.
\\

\noindent $\-$  \small \texttt{Elisabot: How long ago was this picture taken?}\\
$\-$ $>$ \texttt{It was last year}\\
$\-$ \texttt{Elisabot : it was a long time ago. What kind of party is that?}\\
$\-$ $>$ \texttt{It is my retirement party}\\
$\-$ \texttt{Elisabot : it is not too fast. Why are they holding you up?}\\
$\-$ $>$ \texttt{Because we were really close friends}\\
$\-$ \texttt{Elisabot : we can t do that. What is the man looking at? }\\
$\-$ $>$ \texttt{There is no man}\\
$\-$ \texttt{Elisabot : i m sorry. Let's continue with another image, tap /CHANGE, or if you want to leave tap /EXIT }\\

\section{Usability study}

As the target users of this work are older adults with dementia, we decide to evaluate our system with the help of two patients diagnosed of mild cognitive impairment of \textit{Hospital Consorci de Terrassa} and two healthy adults over 60. In this section, we present the built user interface and the feedback obtained from the patients.

\subsection{User interface}
\label{sec:telegram}
 
\begin{figure}
    \centering
    \includegraphics[width=0.6\linewidth]{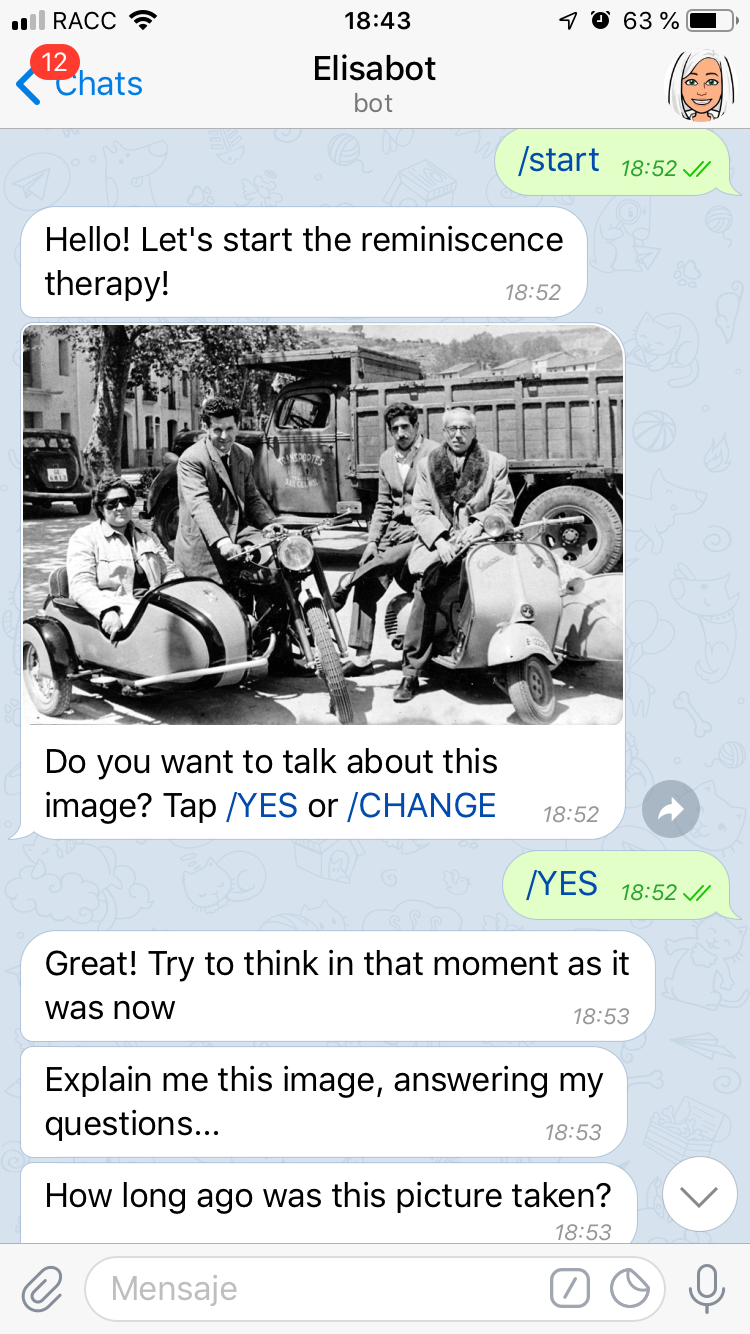}
    \caption{Elisabot interface}
    \label{fig:telegram}
 \end{figure}

We developed a user interface for Elisabot with Telegram, an instant messaging application available for smartphones or computers. 
Telegram is only the interface for the code running in the server.  The bot is executed via an HTTP-request to the API. Users can start a conversation with Elisabot by typing  \texttt{@TherapistElisabot} in the searcher and executing the command \texttt{/start}, as can be seen in the Figure \ref{fig:telegram}. We add more commands like \texttt{/change}, \texttt{/yes} and \texttt{/exit} to enable more functionalities. The commands can be executed either by tapping on the linked text or typing them.

 {}

\subsection{Feedback from patients}

We designed a usability study where users (males and females older than 60 years old) interacted with the system, with the help of a doctor and one of the authors. The purpose was to study the acceptability and feasibility of the system with patients of mild cognitive impairment. We could not do the experiment with more patients as no more patients volunteered for the experiment. The sessions lasted 30 minutes and were carried out by using a laptop computer connected to Telegram. At the end of the session, we administrated a survey to ask participants the following questions about their assessment of Elisabot:

\begin{itemize}
\item Did you like it?
\item Did you find it engaging?
\item How difficult have you found it?
\end{itemize}

Responses were given on a five-point scale ranging from \textit{strongly disagree} (1) to \textit{strongly agree} (5) and \textit{very easy} (1) to \textit{very difficult} (5). The results were 4.6 for amusing and engaging and 2.6 for difficulty. Healthy users found it very easy to use (1/5) and even a bit silly, because of some of the generated questions and comments. Nevertheless, users with mild cognitive impairment found it engaging (5/5) and challenging (4/5), because of the effort they had to make to remember the answers for some of the generated questions.  All the users had in common that they enjoyed doing the therapy with Elisabot.

\section{Conclusions}

We presented a dialogue system for handling sessions of 30 minutes of reminiscence therapy. Elisabot, our conversational agent leads the therapy by showing a picture and generating some questions. The goal of the system is to stimulate the memory and communication skills of the users, as well as improve their mood.
Two models were proposed to generate the dialogue system for the reminiscence therapy. A visual question generator composed of a CNN and a LSTM with attention and a sequence-to-sequence model to generate feedback on the user's answers. 

The qualitative results show that our model can generate questions and feedback well formulated grammatically, but in some occasions not appropriate in content. As expected, it has tendency to produce non-specific answers and to loss its consistency in the comments with respect to what it has said before. However, the overall usability evaluation of the system by users with mild cognitive impairment shows that they found the session very entertaining and challenging. Though, we see that for the proper performance of the therapy is essential a person to support the user to help him/her remember the experiences that are being asked.

\bibliographystyle{ACM-Reference-Format}
\balance 
\bibliography{sample-base}

\appendix

\end{document}